\documentclass[twocolumn,prb,superscriptaddress,aps,showpacs]{revtex4}

\usepackage{graphicx}
\usepackage{dcolumn}
\usepackage{amsmath}
\usepackage{tikz}
\usepackage{subfigure}

\begin{document}
\title{Neutron Scattering Study of Magnetic Anisotropy in a Tetragonal Anti-ferromagnet Bi$_2$CuO$_4$}
\date{\today}
\author{Bo Yuan}
\affiliation{Department of Physics, University of Toronto, Toronto, Ontario, Canada, M5S 1A7}
\author{Nicholas P. Butch}
\affiliation{NIST Center for Neutron Research, National Institute of Standards and Technology, Gaithersburg, Maryland 20899, USA}
\author{Guangyong Xu}
\affiliation{NIST Center for Neutron Research, National Institute of Standards and Technology, Gaithersburg, Maryland 20899, USA}
\author{Barry Winn}
\affiliation{Neutron Scattering Division, Oak Ridge National Laboratory, Oak Ridge, TN, 37831, USA}
\author{J. P. Clancy}
\affiliation{Department of Physics and Astronomy, McMaster University, Hamilton, ON L8S 4M1 Canada}
\author{Young-June Kim}
\affiliation{Department of Physics, University of Toronto, Toronto, Ontario, Canada, M5S 1A7}
\begin{abstract}

We present a comprehensive study of magnon excitations in the tetragonal easy-plane anti-ferromagnet Bi$_2$CuO$_4$ using inelastic neutron scattering and spin wave analyses. The nature of low energy magnons, and hence the anisotropy in this material, has been controversial. We show unambiguously that the low energy magnon spectrum consists of a gapped and a gapless mode, which we attribute to out-of-plane and in-plane spin fluctuations, respectively. We modelled the observed magnon spectrum using linear spin wave analysis of a minimal anisotropic spin model motivated by the lattice symmetry. By studying the magnetic field dependence of the (1, 0, 0) Bragg peak intensity and the in-plane magnon intensity, we observed a spin-flop transition in the $ab$ plane at $\sim0.4$~T which directly indicates the existence of a small in-plane anisotropy that is classically forbidden. It is only by taking into account magnon zero-point fluctuations beyond the linear spin wave approximation, we could explain this in-plane anisotropy and its magnitude, the latter of which is deduced from critical field of the spin-flop transition. The microscopic origins of the observed anisotropic interactions are also discussed. We found that our data is inconsistent with a large Dzyaloshinskii-Moriya interaction, which suggests a potential departure of Bi$_2$CuO$_4$ from the conventional theories of magnetic anisotropy for other cuprates.
\end{abstract}
\maketitle
\section{Introduction}
One of the central topics in magnetism research is the understanding of anisotropic interactions between spins on a microscopic level. Since the full spin rotational symmetry of the magnetic Hamiltonian is broken by these interactions, their elucidation is essential for the description of a material's ground state and its low energy excitations. For example, despite having much smaller magnitudes than the isotropic, or Heisenberg interactions, these anisotropic interactions are responsible for giving rise to a bulk magnetic anisotropy energy (MAE) which determines the ordering direction in an ordered magnet, and providing magneto-elastic/magneto-electric coupling in certain multiferroics\cite{Sergienko2006}. In recent years, materials with anisotropic interactions comparable or even larger than the Heisenberg interactions are beginning to attract much attention\cite{Khaliulin2009}. Dominant anisotropic terms have been shown to modify the magnetism dramatically and give rise to many exotic phenomena in these materials, such as spin liquid phase with topological order \cite{Hermanns2018, Takagi2019, Savary_2016} and topological excitations\cite{Owerre_2016,Chen2018,Aguilera2020}. These new discoveries provide additional impetus to study the anisotropic interactions in different materials.   

In general, the anisotropic interactions arise from the spin-orbit coupling (SOC), and their strength scale with that of the SOC. Although they can be understood qualitatively by examining the local symmetry of the interacting magnetic ions, the magnitude of each symmetry allowed term can only be obtained through the full electronic Hamiltonian including SOC, crystal electric field (CEF), Coulomb interaction ($U_0$) and hopping ($t$). So far, the anisotropic interactions are best understood in $3d$ transition metal materials where SOC is much smaller than the other energy scales, and can therefore be treated perturbatively. In particular, systems where the magnetic ion contains only a single hole in a non-degenerate orbital have been most intensively studied due to its simplicity and direct applicability to high $\mathrm{T_c}$ cuprates. The anisotropic interactions in such systems take the form of an exchange anisotropy between two $S=\frac{1}{2}$-spins (as opposed to single-ion anisotropy that depends on individual spins), which can be written as a sum of an antisymmetric and a symmetric part:
\begin{align}
\vec{A}\cdot(\vec{S}_1\times\vec{S}_2)+\vec{S}_1^{\mathsf{T}} \mathsf{M}\vec{S}_2, 
\label{general}
\end{align}
where $\vec{A}$ and $\mathsf{M}$ are a vector and a symmetric matrix, respectively. First shown by Moriya\cite{Moriya1960}, and later confirmed by Shekhtman and co-workers\cite{Shekhtman1992, Shekhtman1993}, Eq.~\eqref{general}, when expanded up to second order in SOC, takes the following specific form,   
\begin{align}
\vec{D}\cdot(\vec{S}_1\times\vec{S}_2)+\frac{|\vec{D}|^2}{4J}\left[(\hat{d}\cdot\vec{S}_1)(\hat{d}\cdot\vec{S}_2)-\vec{S}_1\cdot\vec{S}_2\right].
\label{DMmodel}
\end{align}
In Eq.~\eqref{DMmodel}, the magnitude of $\vec{D}$ is given approximately by $|\vec{D}|\sim\frac{\lambda}{\epsilon} J$, where $\lambda$, $\epsilon$ and $J$ denote the SOC, CEF and Heisenberg superexchange interaction, respectively. The two terms of Eq.~\eqref{DMmodel}, linear and quadratic in SOC, are often referred to as the Dzyaloshinskii-Moriya (DM) and symmetric anisotropic (SA) interactions. Corrections to Eq.~\eqref{DMmodel} have been investigated by Yildirim $\it{et\,al.}$ in a realistic model\cite{Yildirim1995}, where they showed that Eq.~\eqref{DMmodel} is actually valid up to all orders of SOC, as long as the Coulomb exchange is zero, and the Coulomb interaction is independent of orbitals. The Coulomb exchange, $K$, and parts of the Coulomb interaction depending on orbital occupancy, $\Delta U$, generates corrections to both the DM and SA interactions in Eq.~\eqref{DMmodel} that are an order $\sim\mathcal{O}(\frac{K}{U_0})$ and $\sim\mathcal{O}(\frac{\Delta U}{U_0})$ smaller than the leading terms in Eq.~\eqref{DMmodel}. Such corrections become important when $\vec{D}$ vanishes due to inversion symmetry. The general theory, including both Eq.~\eqref{DMmodel}, and the corrections due to $K$ and $\Delta U$, microscopically explains the magnetic anisotropy in a wide range of tetragonal\cite{Chou1997, Harris2001, Katsumata_2001, Greven1995, Shamoto1993, Yildirim1994_1} and orthorhombic\cite{Peters1988,Keimer1993} Cu$^{2+}$ oxides with a simple Cu-O-Cu bond geometry.

\begin{figure}[!tb]
	\centering
	\includegraphics[width=0.5\textwidth]{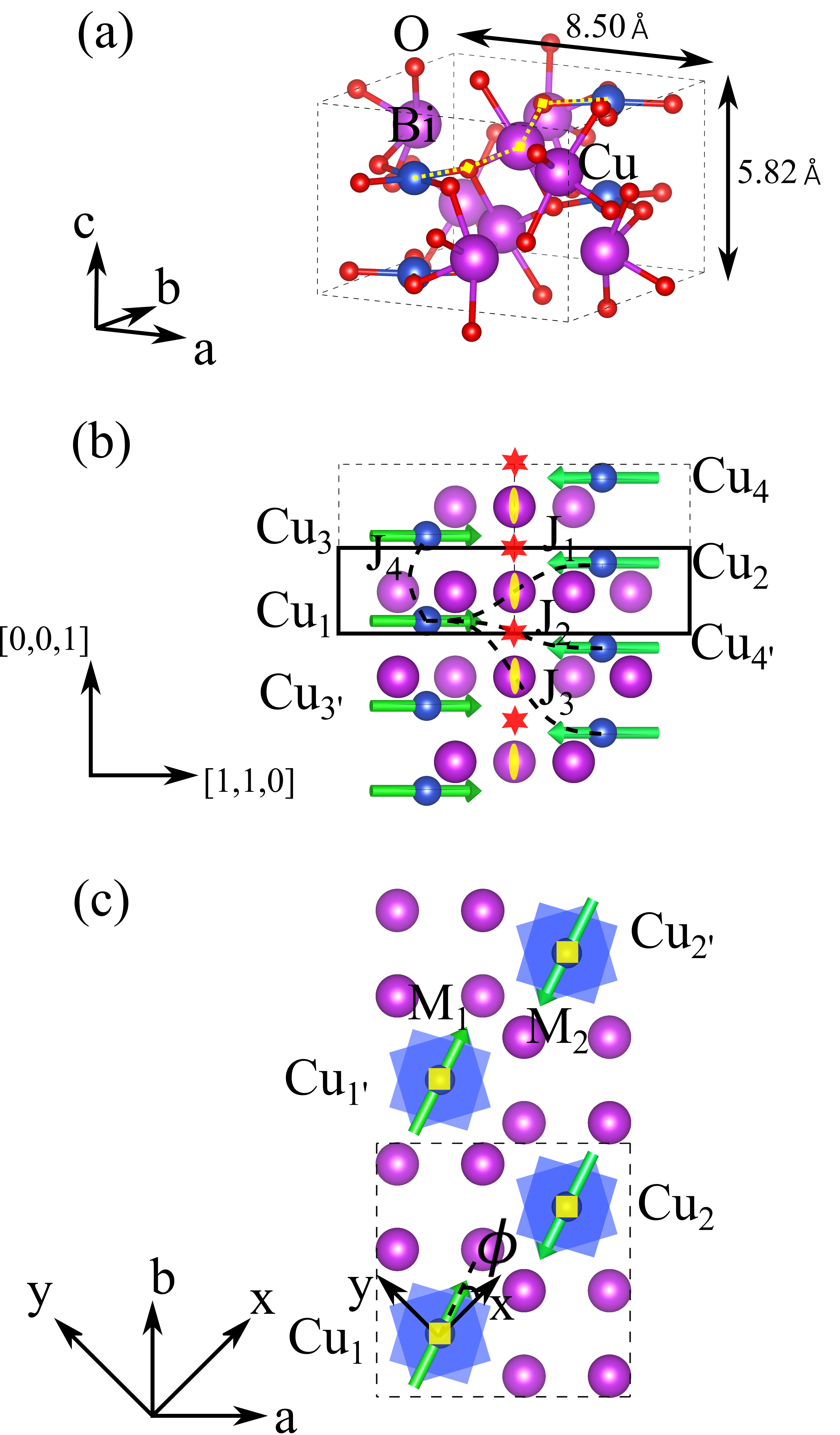}
	\caption{(a) Crystal structure of Bi$_2$CuO$_4$. Oxygen, bismuth and copper ions are shown as red, purple and blue spheres. (b, c) Projection of the Bi$_2$CuO$_4$ structure along the (b) [1, -1, 0] and (c) [0, 0, 1] directions. Oxygen atoms have been omitted in (b) and (c). The local spin axes, $\hat{x}$ and $\hat{y}$ defined in Section \ref{Calculation} are shown in (c). Cu spins are denoted by green arrows, which make an angle $\phi$ with respect to $\hat{x}$. The two sub-lattice magnetizations are indicated by $\vec{M}_1$ and $\vec{M}_2$. Dashed and solid line boxes denote the structural and magnetic unit cell, respectively.  $J_1$, $J_2$, $J_3$ and $J_4$ are the superexchange interactions used in the spin wave calculation.  In (b), the two-fold rotation and inversion symmetries are denoted by yellow ovals and red stars, respectively. In (c), yellow squares denote four-fold rotational symmetry around the chains of Cu ions. }
	\label{structure}	
\end{figure}

A natural next question is the form of the anisotropic interaction when the geometry of the exchange path is more complex. In particular, it is unclear whether the general theory with a dominant anisotropic term given by Eq.~\eqref{DMmodel} still holds when the exchange route is more complicated than a simple Cu-O-Cu bond geometry. One example studied is Bi$_2$CuO$_4$, with a tetragonal lattice structure shown in Fig.~\ref{structure}. The magnetic sub-lattice consists of chains of Cu$^{2+}$ ions arranged in a square lattice. Below $\mathrm{T_N}\sim$50~K, the Cu-moments acquire a $C$-type anti-ferromagnetic order consisting of ferromagnetic chains that are anti-ferromagnetically arranged\cite{bco_powder1,bco_powder2,bco_powder3,yamada1991}. The ordered moments lie in the $ab$ plane\cite{Zhao2017}, suggesting the existence of easy-plane anisotropy similar to other tetragonal cuprates with simpler structure. As shown in Fig.~\ref{structure}, the geometry of the superexchange path between any two Cu$^{2+}$ ions in Bi$_2$CuO$_4$ is very complex. An example is the Cu-O-Bi-O-Cu path indicated by yellow dotted line in Fig.~\ref{structure}(a), where the  Cu-O-Bi and O-Bi-O angles are 109$^\circ$ and 88$^\circ$, respectively. This makes it difficult to theoretically determine the magnetic anisotropy. Experimentally, many studies have been carried out to elucidate the form and size of the anisotropy terms in this material by examining its magnetic excitations. However, the results obtained so far have been inconclusive. Although early inelastic neutron scattering (INS) studies agree on the overall magnon dispersion in Bi$_2$CuO$_4$, they disagree on the nature of the low energy magnons. The first INS study carried out by Ain $\it{et\,al.}$ observed only one low energy magnon mode with a gap of $\sim$~2~meV\cite{Ain1993}. This is in apparent contradiction with the observation of an in-plane ordered moment, which should give two distinct magnon modes polarized in and out of the easy plane. We will henceforth refer to the two modes as `in-plane' and `out-of-plane' magnons. (Note that `in-plane' and `out-of-plane' refer to the polarizations of spin fluctuations, NOT propagation wave vectors.) Subsequent INS by Roessli $\it{et\,al.}$ found two modes, with a gap of $\sim$~4~meV and $\sim$~2~meV\cite{Roessli1993}. Later polarized INS experiment by the same authors reported different gap sizes of $\sim$~0.5~meV and $\sim$~2~meV, which they attributed to in-plane and out-of-plane spin fluctuations\cite{Roessli1997}, respectively. Observation of such a large in-plane magnon gap in an easy-plane magnet with tetragonal lattice symmetry is entirely unexpected by symmetry (see Section~\ref{symmetryargument}), which motivated early theoretical work that introduced a highly unconventional four-spin interaction\cite{Petrakovskii1994, Roessli1993}. Lastly, all the INS results to date contradicted the antiferromagnetic resonance (AFMR) results\cite{Hitoshi1992,Ohta1998}, which reported a gapped and a gapless magnon mode.

To reconcile the controversies in earlier studies, and further the understanding of anisotropic interactions in Bi$_2$CuO$_4$, we carried out new high resolution INS studies of magnetic excitations in a Bi$_2$CuO$_4$ single crystal. Using thermal neutron with an intermediate energy resolution, we mapped out the full magnon spectrum. The overall magnon dispersion relation observed in our study is qualitatively consistent with earlier results\cite{Ain1993, Roessli1993}. However, using cold neutron with much higher resolution than all previous INS studies, we show that the low energy spectra consist of two modes, one is gapless and one has a $\sim$~2~meV gap, that can be attributed to in-plane and out-of-plane magnons, respectively. Our results therefore confirm the AFMR results and resolve the controversies regarding the nature of the low energy excitations. We also carried out INS in the presence of a magnetic field applied along the (0, 1, 0) direction. By studying field dependence of the (1, 0, 0) magnetic Bragg peak intensity, and that of the gapless mode, we show that there is a spin-flop transition within the $ab$ plane at $\sim$~0.4~T, which directly indicates the existence of an in-plane MAE that selects the direction of ordered moments in the $ab$ plane. Using a symmetry argument, we show that such an MAE is forbidden on a mean-field level in Bi$_2$CuO$_4$, and therefore could only be explained by considering corrections to the ground state energy due to quantum fluctuations, a phenomena known as quantum order by disorder\cite{Rau2018}. We could explain the observed magnon dispersion by carrying out linear spin wave analysis on a model containing Heisenberg interaction and a symmetry allowed exchange anisotropy of the form, $\alpha_\parallel S_{1,x} S_{2,x}-\alpha_\parallel S_{1,y} S_{2,y}+\alpha_\perp S_{1,z} S_{2,z}$, where $2\alpha_\parallel\sim \alpha_\perp\sim$ 0.01. Beyond the linear spin wave approximation, we considered corrections to the ground state energy due to magnon zero-point fluctuations. This predicts an in-plane MAE that quantitatively explains the critical field of the spin-flop transition.          

\section{Experimental Details}
Bi$_2$CuO$_4$ single crystal (4.35~g) used for neutron scattering measurements was grown using the floating zone technique\cite{Zhao2017}. Full magnon spectrum of Bi$_2$CuO$_4$ was mapped out using the HYSPEC time-of-flight (TOF) spectrometer at the Spallation Neutron Source (SNS) at Oak Ridge National Laboratory. An incident energy of $\mathrm{E_i}= 25$~meV was used to give an energy resolution of $\sim$ 1~meV at zero energy transfer. Field dependence of magnon below $\sim$ 4~meV was studied using the Disk Chopper Spectrometer (DCS) and Spin Polarized Inelastic Neutron Spectrometer (SPINS) at the NIST Center for Neutron Research (NCNR). Two incident energies of $\mathrm{E_i}=$ 4.9~meV and $\mathrm{E_i}=$ 2.3~meV were used for TOF measurement at DCS, which gave an energy resolution of $\sim$ 0.18~meV and $\sim$ 0.06~meV at the elastic line, respectively. The triple-axis measurement at SPINS was performed using a fixed final energy $\mathrm{E_f}=$ 5~meV. A vertically focussing pyrolitic graphite (PG) monochrometer, flat PG analyzer and a Be filter were used to select incident and final energies at SPINS. A collimation setting of guide-open-80'-open was used to achieve an energy resolution of $\sim$ 0.2~meV at the elastic line. For all measurements, the crystal was aligned with (H, 0, L) in the scattering plane. The alignment was carried out at the McMaster Alignment Diffractometer (MAD) prior to the INS experiments. A 10~T vertical field superconducting magnet was used for measurements at DCS and SPINS to apply a field along the (0, 1, 0) direction. The sample temperature was kept at $\sim$~1.5~K for all measurements. 

\section{Experimental Results}
\begin{figure}[!tb]
	\centering
	\includegraphics[width=0.5\textwidth]{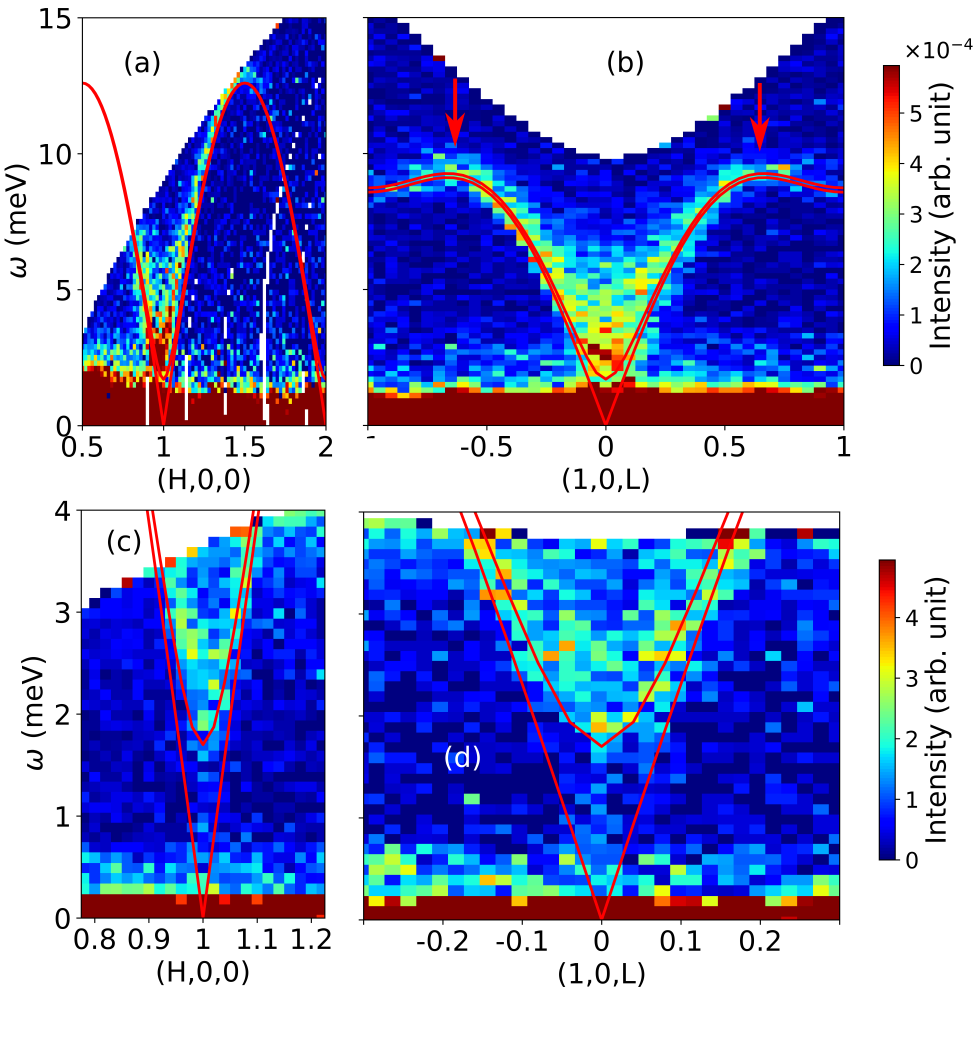}
	\caption{(a,b) INS spectra as a function of energy, $\omega$,(y-axis) and momentum transfer (x-axis) along (a) (H, 0, 0) and (b) (1, 0, L). We set $\hbar~=~1$ throughout the paper. The data is obtained at HYSPEC using an incident energy of $\mathrm{E_i}$~=~25~meV. Vertical arrows in (b) denote the positions of local maxima in the dispersion along L. (c,d) High resolution INS spectra along (c) (H, 0, 0) and (d) (1, 0, L) obtained at DCS using an $\mathrm{E_i}=$ 4.7~meV. The intensity scales used to plot the HYSPEC and DCS data are shown to the right of (b) and (d). Red solid lines in all plots are calculated dispersion relations within the XYZ model, as described in the text.}
	\label{zerofield}	
\end{figure}

Full magnon dispersion in Bi$_2$CuO$_4$ mapped out using the HYSPEC TOF spectrometer is shown in Fig.~\ref{zerofield}(a)-(b). As shown in Fig.~\ref{zerofield}(a), the energy of the magnon increases monotonically along H, reaching an energy of $\omega=$ 12.6~meV at the magnetic zone boundary at $\mathbf{Q}=$ (1.5, 0, 0). On the other hand, the magnon along L first disperses upward, reaching a maximum energy of $\omega=$ 9.2~meV at $\mathbf{Q}=$ (1, 0, $\pm$0.7) (denoted by vertical arrows in Fig.~\ref{zerofield}(b)), before curving down to an energy of $\omega=$ 8.6~meV at the zone boundary with $\mathbf{Q}=$ (1, 0, $\pm$1). These features are in qualitative agreement with early INS results reported by other authors\cite{Ain1993, Roessli1993}. Results of high resolution measurement of magnon excitations below $\sim$ 4~meV from the DCS experiment are shown in Fig.~\ref{zerofield}(c) and Fig.~\ref{zerofield}(d) for momentum transfer along H and L, respectively. Clearly, a magnon gap of $\sim$ 2~meV can be resolved with the high resolution data, consistent with Ref.~[\onlinecite{Ain1993, Roessli1993,Roessli1997}]. 

\begin{figure*}[!tb]
	\centering
	\includegraphics[width=1\textwidth]{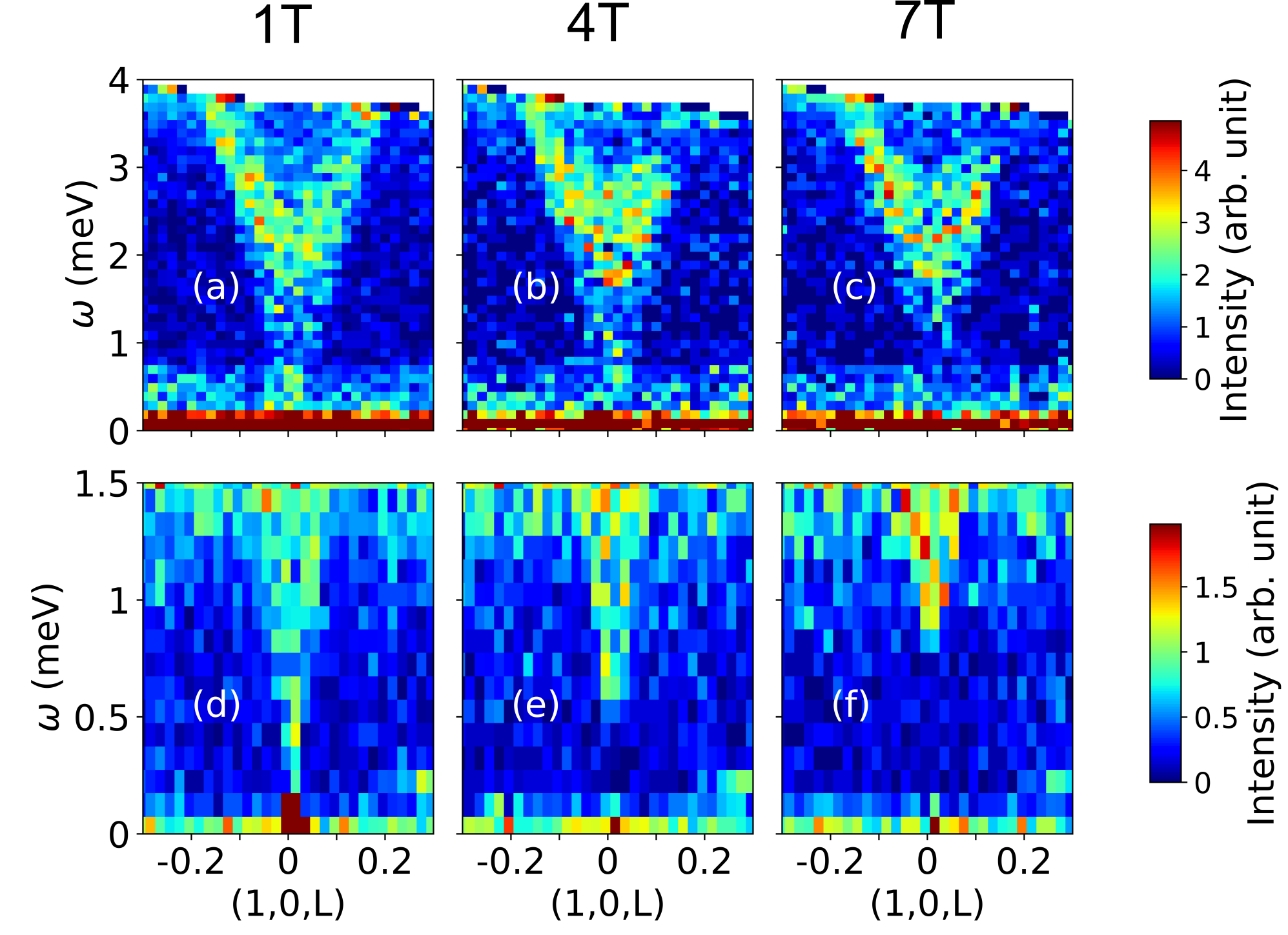}
	\caption{ (a,b,c) INS spectra along (1, 0, L) obtained at DCS using $\mathrm{E_i}=$ 4.7~meV for different fields along (0, 1, 0). Magnitude of the field is given on top of each figure. (d,e,f) Same as (a,b,c) but obtained using $\mathrm{E_i}=$ 2.3~meV which gives higher energy resolution. Intensity scales used for the two $\mathrm{E_i}$'s are shown on the right.}
	\label{fieldep}	
\end{figure*}

However, even at zero field, there seems to be a small but non-zero intensity below the gapped mode in Fig.~\ref{zerofield}(c) and Fig.~\ref{zerofield}(d). As shown in Fig.~\ref{fieldep}(a), the intensity below the $\sim$ 2~meV gap can be enhanced by applying a small magnetic field (1~T) along the (0, 1, 0) direction. The shape of its dispersion clearly shows that it originates from another mode with an acoustic-like dispersion. The acoustic mode merges with the gapped mode away from the magnetic zone center, making them indistinguishable at energy transfer larger than $\sim$ 2~meV. To determine the presence of any small gap in this acoustic-like mode, we carried out measurements using an $\mathrm{E_i}=$ 2.3~meV with higher resolution. Results in Fig.~\ref{fieldep}(d) clearly show that this mode is gapless within an experimental resolution of $\sim$ 0.06~meV. 

Given the highly symmetric crystal structure of Bi$_2$CuO$_4$, one expects the ordered moments to be able to rotate freely within the $ab$ plane, giving rise to a gapless magnon mode due to the in-plane spin fluctuation. On the other hand, an easy-plane anisotropy is expected from the observation that the ordered moments lie in the $ab$ plane, implying that the magnon mode due to out-of-plane spin fluctuation must be gapped. These arguments allow us to assign the acoustic and gapped magnon mode observed in our data to in-plane and out-of-plane spin fluctuations, respectively.  

As shown in Fig.~\ref{fieldep}(e)-(f) and Fig.~\ref{fieldep}(b)-(c), applying higher fields gaps out the acoustic mode while leaving the gapped mode unchanged. Since a magnetic field along (0, 1, 0) breaks the spin rotational symmetry within the easy-plane, in-plane spin fluctuation should acquire a gap roughly equal to the Zeeman energy, or $g\mu_B H$. Using $g\approx 2$ for $\mathrm{Cu^{2+}}$[\onlinecite{Ohta1998}], we estimate the in-plane magnon gap
to be $\sim$ 0.5~meV and $\sim$ 0.8~meV at 4~T and 7~T, in quantitative agreement with our data in Fig.~\ref{fieldep}(e) and Fig.~\ref{fieldep}(f). On the other hand, the out-of-plane magnon gap is unaffected by the field as it is much larger than the Zeeman energy for the range of fields used in our experiment.

\begin{figure}[!tb]
	\centering
	\includegraphics[width=0.5\textwidth]{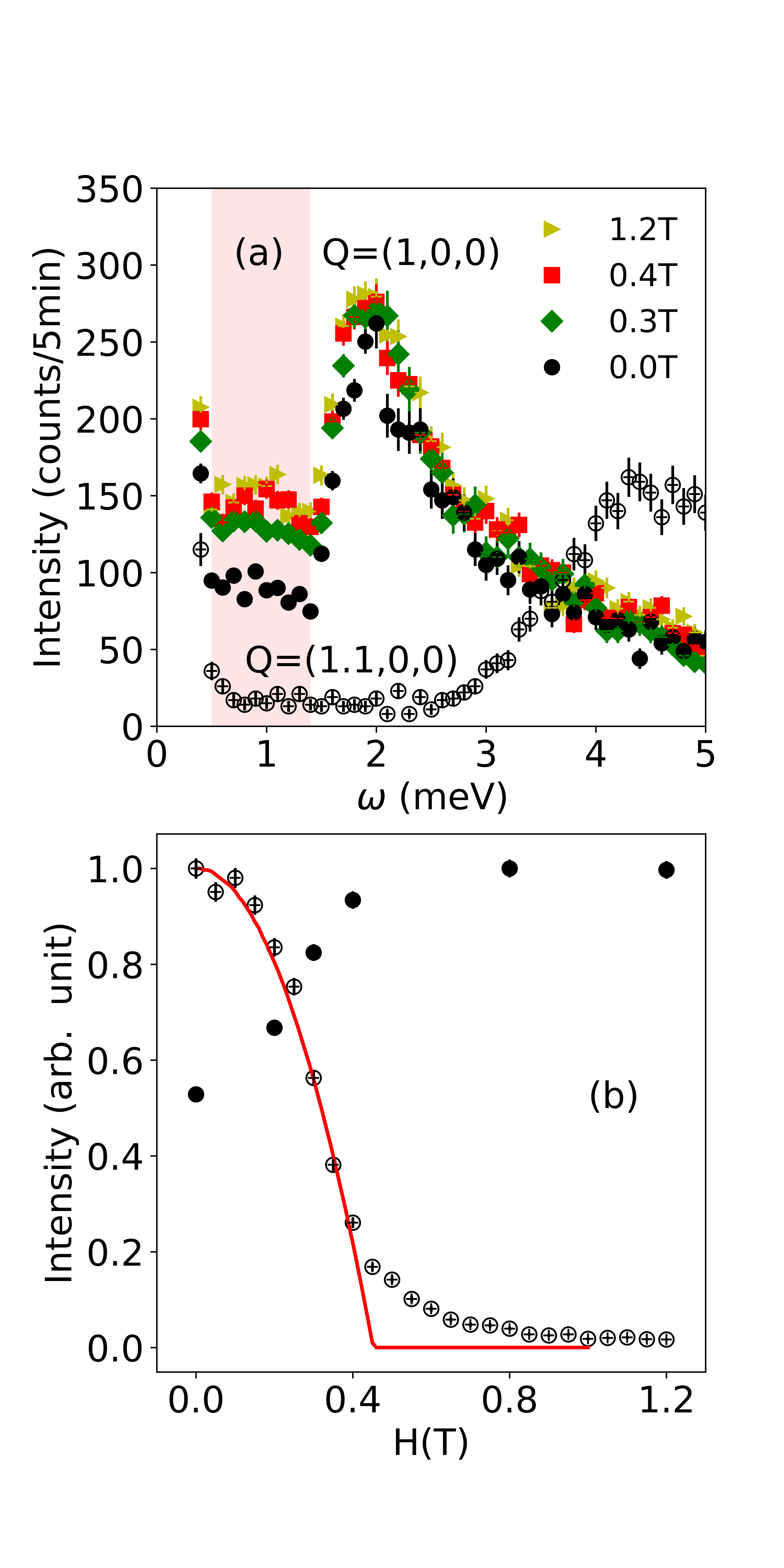}
	\caption{(a) Energy scan at constant $\bf{Q}=$ (1, 0, 0) for different fields (0~T-1.2~T) applied along (0, 1, 0). Data shown in open circle is an energy scan at $\bf{Q}$ = (1.1, 0, 0) at zero field, which is used as non-magnetic background for energy transfer $0.5~\mathrm{meV}\lesssim\omega\lesssim~2.5~\mathrm{meV}$. The data is obtained at SPINS with a fixed energy $\mathrm{E_f}=$ 5~meV. (b) Field dependence of the (1, 0, 0) magnetic Bragg peak intensity (open circle) and intensity of the gapless mode (solid circle). Intensity of the gapless mode is obtained by integrating the constant $\bf{Q}$ scan at (1, 0, 0) from 0.5~meV to 1.5~meV (shaded region in (a)) after subtracting the background at (1.1, 0, 0) within the same energy range. Intensities of the magnetic Bragg peak and the gapless mode have been normalized with respect to the values at 0~T and 1.2~T respectively. Solid line is the calculated (1, 0, 0) Bragg peak intensity as described in Section~\ref{spinflopsection}.}
	\label{SPINSdata}	
\end{figure}

To study the intensity change of the two modes at $H\lesssim$ 1~T in greater detail, an energy scan at constant $\mathbf{Q}=$ (1, 0, 0) was carried out for different fields below 1.2~T as shown in Fig.~\ref{SPINSdata}(a). The peak at $\sim$ 2~meV in all energy scans corresponds to the gapped out-of-plane magnon mode shown in Fig.~\ref{fieldep}(a)-(c). Clearly, the intensity of this mode is almost independent of applied field. We also plot the energy scan at $\bf{Q}$ = (1.1, 0, 0) obtained at zero field in Fig.~\ref{SPINSdata}(a) with open circles. Since the magnon has a fairly steep dispersion,  it has dispersed to higher energy ($\omega>$ 3~meV) at this $\bf{Q}$ and the small residual intensity for energy transfers 0.5~meV $\lesssim\omega\lesssim$ 3~meV can be taken as the background. Compared to the $\bf{Q}$ = (1.1, 0, 0) data, the scan at $\bf{Q}$ = (1, 0, 0) at zero field clearly shows additional inelastic intensity below the $\sim$ 2~meV peak that extends down to elastic region. This intensity comes from the gapless mode identified from our DCS data (Fig.~\ref{fieldep}(d)). As shown in Fig.~\ref{SPINSdata}(a), this intensity increases with field. To obtain a quantitative measure of the in-plane magnon mode intensity, the background at $\bf{Q}$ = (1.1, 0, 0) was first subtracted from the energy scan at $\bf{Q}$ = (1, 0, 0), and then the intensity from 0.5~meV to 1.5~meV (shaded region in Fig.~\ref{SPINSdata}(a)) was integrated. The integrated intensity is plotted as a function of applied field strength in Fig.~\ref{SPINSdata}(b) (solid circle). One can observe that the in-plane mode intensity in this energy range almost doubles from 0~T to $\sim$ 0.4~T and then stays the same beyond this field. Also shown in Fig.~\ref{SPINSdata}(b) is the intensity of the magnetic Bragg peak at $\bf{Q}$ = (1, 0, 0) represented by open circles, which is completely suppressed as the intensity of the in-plane mode reaches its maximum. Both the intensity of the in-plane mode and the (1, 0, 0) magnetic Bragg peak intensity changes at $\sim0.4$~T. This coincides with the meta-magnetic transition observed in bulk magnetization studies, which was attributed to a spin-flop transition\cite{yamada1991}. As we will explain in Section~\ref{Calculation}, the intensity change is quantitatively consistent with re-orientation of spins due to a spin flop transition within the $ab$ plane. 

In the absence of any magnetic anisotropy in the $ab$ plane, spins can rotate freely and respond to even an infinitesimal field. Observation of a spin-flop transition at a finite field therefore directly indicates that in addition to the easy-plane anisotropy, there exists a finite in-plane magnetic anisotropy energy (MAE).

\section{Spin Wave Analysis}\label{Calculation}
\subsection{Heisenberg Model}
\begin{figure}[!tb]
	\centering
\includegraphics[width=0.5\textwidth]{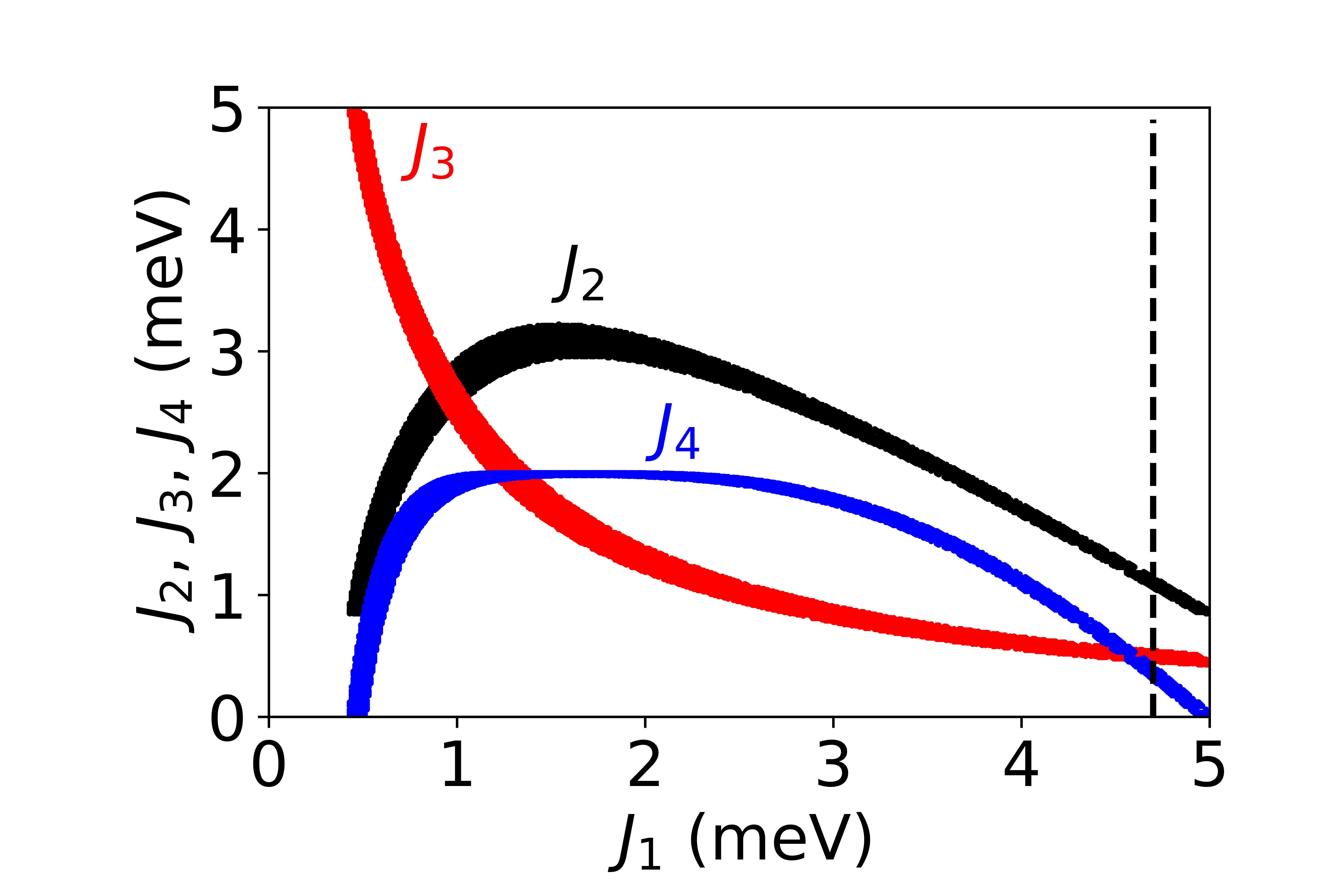}
	\caption{Allowed $J_2$, $J_3$ and $J_4$ for a given $J_1$. Thickness of the lines denotes the range of allowed values for each parameter, which reflects uncertainties in the constraints used to determine these parameters. The values of $J_1$ to $J_4$  used in our spin wave calculation is indicated by dashed line.}
	\label{Heisenbergparameters}	
\end{figure}

In this subsection, we provide a quantitative description of our data within linear spin wave theory (See Appendix B for details). The dominant superexchange interactions in Bi$_2$CuO$_4$,  labelled by $J_1$, $J_2$, $J_3$ and $J_4$ in Fig.~\ref{structure}(b), were proposed in previous studies, and corroborated by an $\it{ab\,initio}$ calculation\cite{Janson2007}. Since the exchange anisotropy in each bond is expected to be much smaller than the isotropic part, we first determine the magnitude of the latter by comparing our data to a Heisenberg model, which greatly simplifies the calculation. As we show by symmetry analysis in Appendix A, all interactions centered at $\mathrm{Cu_3}$ can be obtained from those centered at $\mathrm{Cu_1}$ by a combination of inversion and 2-fold rotation (These symmetry operations have been indicated in Fig.~\ref{structure}). Since a Heisenberg interaction of the form $J\vec{S}_i\cdot\vec{S}_j$ is left invariant by these symmetry operations, interactions centered at $\mathrm{Cu_3}$ and $\mathrm{Cu_1}$ are therefore identical within the Heisenberg model, making it invariant under a translation by $\it{half}$ of the structural unit cell along the $c$ direction. The primitive magnetic unit cell within the Heisenberg model (solid line in Fig.~\ref{structure}(b)) is therefore half of the structural unit cell (dashed line in Fig.~\ref{structure}(b)), and consists of only two Cu ions. Linear spin wave analysis therefore predicts two magnon modes, whose energies are degenerate and are given by the following simple analytic expression ,
\begin{align}
\begin{split}
\omega(\mathbf{Q})&=\biggr\{\left[J_4(\cos(\pi L)-1)+2J_1+2J_2+2J_3\right]^2\\
&-\left|\frac{1}{2}\left[A_{-\mathbf{Q}}+B_{-\mathbf{Q}}\right]\left[J_1+J_2e^{-i\pi L}+J_3e^{-i2\pi L}\right]\right|^2\biggr\}^{\frac{1}{2}}, 
\end{split}
\label{omegaHeisenberg}
\end{align}
where $A_\mathbf{Q}=1+e^{i2\pi(H+K)}$, $B_\mathbf{Q}=e^{i2\pi H}+e^{i2\pi K}$ and $\mathbf{Q}$ is given in the reciprocal lattice unit of the structural unit cell. Rather than fitting to the full magnon spectrum as done previously by other authors\citep{Ain1993, Roessli1993}, we note that the exchange parameters can already be determined by three independent quantities: magnon energies at $\mathbf{Q}=(1.5,~0,~0)$ and $\mathbf{Q}=(1,~0,~1)$, as well as energy and $\mathbf{Q}$ of the local maximum in the L-dispersion shown in Fig.~\ref{zerofield}(b). The magnon zone boundary energies in Fig.~\ref{zerofield}(a) and Fig.~\ref{zerofield}(b) are determined to be $\omega=$~12.6(2)~meV and $\omega=$~8.6(3)~meV, respectively. However, location of the local maximum in Fig.~\ref{zerofield}(b) can only be roughly determined to be at $\omega$~=~9-9.5~meV and L~=~0.65-0.75. In Fig.~\ref{Heisenbergparameters}, we show the allowed parameters obtained from these constraints. For a given $J_1$, $J_2$ to $J_4$ are more or less fixed. However, $J_1$ itself can vary from $\sim$0.5~meV to $\sim$5~meV, implying that there is no unique set of `best fit' parameters as suggested by previous INS studies. We have checked that all these parameters give almost identical magnon dispersions that describe our data equally well. Motivated by results from the $\it{ab\,initio}$ calculations showing $J_1\gg J_2, J_3, J_4$, we use the representative set of parameters: $J_1=4.7$~meV, $J_2$~=~1.1~meV, $J_3$~=~0.5~meV and $J_4$~=~0.36~meV, where we have fixed the value of $J_1$ to that estimated from the tight-binding and LSDA+U calculation in Ref.~[\onlinecite{Janson2007}]. These values are indicated by the dashed line in Fig.~\ref{Heisenbergparameters}.

\subsection{Exchange Anisotropy} 
To capture the magnetic anisotropy in Bi$_2$CuO$_4$, we have to go beyond the simple Heisenberg model and include anisotropic terms. For a given bond, the exchange anisotropy should roughly scale with the isotropic part, both of which are proportional to $\frac{t^2}{U_0}$ in the large $U_0$ limit. We will therefore only consider the exchange anisotropy for the dominant exchange path, $J_1$. The simplest symmetry allowed anisotropic exchange beyond the Heisenberg model (see Appendix A) is given by $J_{xx}^1 S_{x,1}S_{x,2}+J_{yy}^1 S_{y,1}S_{y,2}+J_{zz}^1 S_{z,1}S_{z,2}$. We will call this an XYZ model. Taking this to be the interaction between $\mathrm{Cu_1}$ and $\mathrm{Cu_2}$, 4-fold symmetry implies that the interaction between $\mathrm{Cu_{1^\prime}}$ and $\mathrm{Cu_2}$ (Fig.~\ref{structure}(c)) is $J_{yy}^1 S_{x,1^\prime}S_{x,2}+J_{xx}^1 S_{y,1^\prime}S_{y,2}+J_{zz}^1 S_{z,1^\prime}S_{z,2}$. In these expressions, we have defined a new coordinate system $xyz$, where $\hat{x}$ and $\hat{y}$ are rotated by 45$^{\circ}$ with respect to the crystallographic $\hat{a}$ and $\hat{b}$ directions, and $\hat{z}\parallel \hat{c}$. These definitions are motivated by the observation that $\hat{y}$ ($\hat{x}$) is the local two-fold axis for a bond directed along $\hat{x}$ ($\hat{y}$). We emphasize that an XYZ model is a minimal model that can quantitatively account for all the observations in our data, notably the existence of both an out-of-plane magnon gap and an in-plane MAE revealed by a spin-flop transition. As we show by symmetry analysis (see Appendix A), other exchange anisotropy such as off-diagonal SA and DM interactions are also allowed for the $\mathrm{Cu_1-Cu_2}$ bond. Although we cannot rule out the former in Bi$_2$CuO$_4$, we will show later that a large DM interaction is inconsistent with the observed magnon spectrum.   
  
A magnetic unit cell half of the structural unit cell can still be used for a spin wave calculation within the XYZ model (See Appendix A). The angle between the sub-lattice magnetization, $\vec{M}_1$, (or equivalently the staggered moment, $\vec{n}=\vec{M_1}-\vec{M_2}$) and the $\hat{x}$-axis is given by $\phi$ as shown in Fig.~\ref{structure}(c) (also in Fig.~\ref{spinflop}(c)). Since $J_{xx}\neq J_{yy}$ breaks the in-plane spin rotational symmetry of the Hamiltonian, the resulting magnon spectrum is expected to have an explicit $\phi$ dependence. Within linear spin wave theory, energies of the two magnon modes are given by

\begin{align}
\begin{split}
\omega_\pm&=\biggr(C_{\mathbf{Q}}^2+|E_{\mathbf{Q}}|^2-|F_{\mathbf{Q}}|^2\\
&\pm \sqrt{4|C_{\mathbf{Q}}E_{\mathbf{Q}}|^2-|\bar{E}_{\mathbf{Q}} F_{\mathbf{Q}}-E_{\mathbf{Q}}\bar{F}_{\mathbf{Q}}|^2}\biggr)^\frac{1}{2},
\end{split} 
\label{XYZ}
\end{align}

where

\begin{align}
\begin{split}
C_{\mathbf{Q}}&=J_4\cos(\pi L)-J_4+(J_{xx}^1+J_{yy}^1)+2J_2+2J_3\\
E_{\mathbf{Q}}&=\frac{1}{4}[-J_{xx}^1(\sin^2(\phi)A_{-\mathbf{Q}}+\cos^2(\phi)B_{-\mathbf{Q}})\\
&-J_{yy}^1(\sin^2(\phi)B_{-\mathbf{Q}}+\cos^2(\phi)A_{-\mathbf{Q}})+J_{zz}^1(A_{-\mathbf{Q}}+B_{-\mathbf{Q}})]\\
F_{\mathbf{Q}}&=\frac{1}{4}[J_{xx}^1(\sin^2(\phi)A_{-\mathbf{Q}}+\cos^2(\phi)B_{-\mathbf{Q}})\\
&+J_{yy}^1(\sin^2(\phi)B_{-\mathbf{Q}}+\cos^2(\phi)A_{-\mathbf{Q}})+J_{zz}^1(A_{-\mathbf{Q}}+B_{-\mathbf{Q}})]\\
&+\frac{1}{2}(A_{-\mathbf{Q}}+B_{-\mathbf{Q}})[J_2e^{-i\pi L}+J_3e^{-i2\pi L}],
\end{split}
\label{XYZparameters}
\end{align}

and $\bar{E}_\mathbf{Q}$ and $\bar{F}_\mathbf{Q}$ denoted their complex conjugates. In the above expressions, $J_2$, $J_3$ and $J_4$ are set to be the values determined in the last subsection. Different components of $J_{\zeta\zeta}^1$ are parametrized by the following: $J_{yy/xx}^1=J_1(1\pm\alpha_\parallel)$, and $J_{zz}^1=J_1(1-\alpha_\perp)$, where the isotropic part, $J_1$, is fixed to be 4.7~meV and $\alpha_\parallel$ and $\alpha_\perp$ are to be determined.

\begin{figure}[!tb]
	\centering
\includegraphics[width=0.5\textwidth]{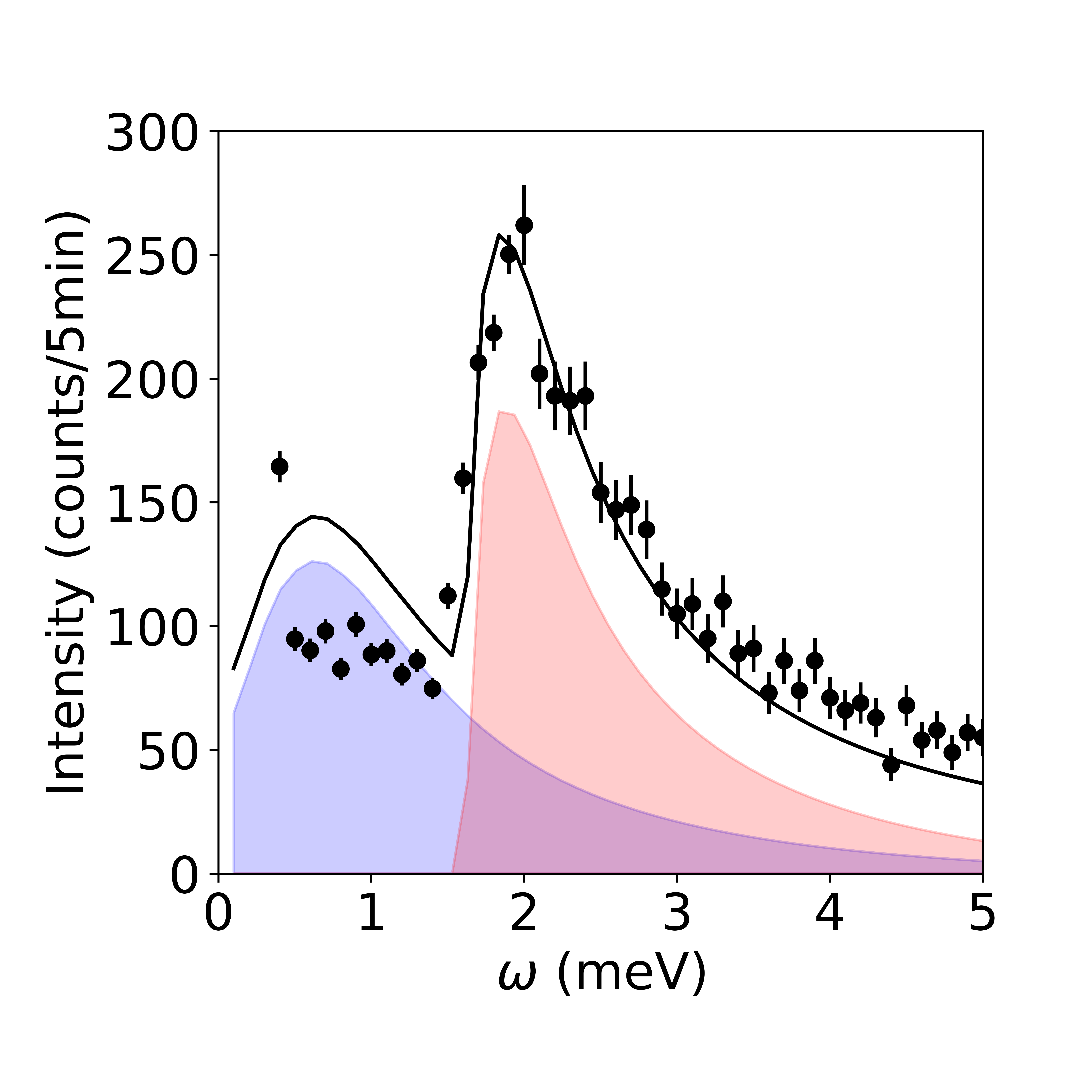}
	\caption{Energy scan at $\mathbf{Q}=$(1,0,0) at 0~T fit to a sum of in-plane and out-of-plane magnon mode as described in the text. Contributions of the two modes are shaded in blue and pink, respectively.}
	\label{cutcalculation}	
\end{figure}

At the magnetic Zone center, $\mathbf{Q}=$ (1, 0, 0), energies of the two modes are $\omega_-=0$ and $\omega_{+}=2\sqrt{(2J_1+2J_2+2J_3)J_1\alpha_\perp}$, respectively, corresponding to in-plane and out-of-plane spin fluctuations observed in our data. Note that the in-plane gap remains zero within the linear spin wave approximation despite a non-zero in-plane exchange anisotropy, $\alpha_\parallel$. To determine $\alpha_\perp$, we fit the zero-field energy scan data shown in Fig.~\ref{SPINSdata}(a) to the following form of dynamical structure factor convolved with the instrumental resolution:
\begin{align}
S(\omega)=\frac{1}{1-\exp(-\omega/k_B T)}\left(\frac{1}{2}\frac{\delta(\omega-\omega_-)}{\omega_-}+\frac{\delta(\omega-\omega_+)}{\omega_+}\right),
\label{S}
\end{align}
plus a constant background. In Eq.~\eqref{S}, $k_B$ is the Boltzmann constant, the first and second term are contributions by the in-plane and out-of-plane spin fluctuations, respectively. The factor of 1/2 for the in-plane spin fluctuation will be explained in Section~\ref{spinflopsection}. As shown in Fig.~\ref{cutcalculation},  a gap of $1.7(2)$~meV, or $\alpha_\perp=0.013(2)$ gives a good description of the data: a peak at $\sim$~2~meV followed by a long tail extending up to $\sim$~5~meV that results from convolving the steep magnon dispersion with the instrumental resolution. The slight discrepancy between the relative intensities of the two modes in our data and the calculation might be due to additional momentum dependent factors in the scattering intensity not captured by Eq.~\eqref{S}.

\subsection{Quantum Order by Disorder}

\subsubsection{Symmetry Argument for Accidental Degeneracy}\label{symmetryargument}
Since the in-plane spin rotational symmetry is absent in the XYZ model, the magnon dispersions given by Eq.~\eqref{XYZ} and Eq.~\eqref{XYZparameters} are expected to depend explicitly on the ordering direction, $\phi$. However, the in-plane magnon gap, $\omega_-$, predicted by linear spin wave theory is exactly 0, independent of $\phi$. We now argue that this somewhat surprising finding is a consequence of the unique crystal symmetry of Bi$_2$CuO$_4$, independent of the underlying microscopic magnetic Hamiltonian. 

We consider a generic ordered state where the magnetic unit cell is the same as the structural unit cell (the magnetic order observed in Bi$_2$CuO$_4$ is a special case where the unit cell of the ordered structure is half of the structural unit cell). Ordered magnetic moments on Cu$_1$-Cu$_4$ are given by $\vec{M}_1$ to $\vec{M}_4$. If we consider only interactions quadratic in spin operators, the mean-field energy of the system is given by $E_\mathrm{MF}=\sum_{i,j}^{\zeta\eta} M_i^\zeta \Gamma_{i,j}^{\zeta\eta} M_j^\eta$ where $i,j=1-4$ denotes the spin index and $\zeta,\eta=x,y,z$ denotes the spin component. The matrix $\Gamma_{i,j}^{\zeta\eta}$ is obtained by summing all exchange interactions. 

Next we note that the lattice is invariant under a rotation by $90^\circ$ about the $c$ axis passing through the chain of Cu atoms. This operation both moves the atoms as well as rotates their spins. However, given the lattice structure of Bi$_2$CuO$_4$, this operation only moves atoms within its own sub-lattice. In other words, Cu$_i$ in one unit cell is moved to the same position in another unit cell. It is therefore equivalent to rigidly rotating the $\vec{M}_i$ by 90$^\circ$ which maps $M_{i,x}\rightarrow M_{i,y}$ , $M_{i,y}\rightarrow -M_{i,x}$ and $M_{i,z}\rightarrow M_{i,z}$ for $i=$1 to 4. This constrains the mean-field energy to be 
\begin{align}
\begin{split}
E_\mathrm{MF}&=\Gamma_{i,j}^{xx}(M_{i,x} M_{j,x}+M_{i,y} M_{j,y})+\Gamma_{i,j}^{zz} M_{i,z} M_{j,z}\\
&+\Gamma_{i,j}^{xy} (M_{i,x} M_{j,y}-M_{i,y} M_{j,x})
\end{split}. 
\label{Emf}
\end{align}

Clearly, $E_\mathrm{MF}$ acquires an accidental in-plane spin rotational symmetry. This implies that all ordering directions within the $ab$ plane are classically degenerate. Moreover, since the in-plane magnon at the zone center corresponds to a rigid rotation of all spins that costs zero energy on a mean-field level, it must remain gapless within linear spin wave approximation. Note that this argument only works because of the simple ferromagnetic chain arrangement of Cu ions in Bi$_2$CuO$_4$, which is unique among the tetragonal cuprates. Similar arguments fail, for example, in Sr$_2$CuO$_2$Cl$_2$ where the adjacent Cu ions along the $c$ direction are displaced diagonally by half of the unit cell, in which case an anisotropic term is allowed in the mean-field energy due to inter-layer coupling\cite{Yildirim1994_1}.

\subsubsection{Magnon Zero-point Fluctuations} 
Although no magnetic anisotropy within the $ab$ plane is expected on a classical level, summing magnon zero-point fluctuations (ZPF) at all $\mathbf{Q}$'s generates a correction to the ground state energy that is explicitly $\phi$ dependent, which contributes to the bulk magnetic anisotropy energy (MAE) of Bi$_2$CuO$_4$. Quantitatively, this correction is given (up to a constant) by\cite{Rau2018}
\begin{align}
E_\mathrm{ZPF}(\phi)=\frac{1}{2}\sum_\mathbf{Q}(\omega_+(\phi) +\omega_-(\phi)).
\label{QZPF}
\end{align}

Carrying out the sum numerically gives an $E_\mathrm{ZPF}(\phi)$ shown in Fig.~\ref{spinflop}(a) that is minimum at $\phi=0$ and maximum at $\phi=45^\circ$. The ordered moments therefore lie along the Cu-Cu bond directions ($\hat{x}$ or $\hat{y}$) at zero field. $E_\mathrm{ZPF}(\phi)$ can be fit very well by $E_\mathrm{ZPF}(\phi)=\Lambda \sin(2\phi)^2$, where $\Lambda$ gives the difference $E_\mathrm{ZPF}(45^\circ)-E_\mathrm{ZPF}(0^\circ)$. $\Lambda$ as a function of $\alpha_\parallel$ is shown in Fig.~\ref{spinflop}(b), which is described very well by a quadratic function. The fact that $E_\mathrm{ZPF}$ is an even function of $\alpha_\parallel$ can be understood by noting that a sign change of $\alpha_\parallel$ is equivalent to interchanging $S_x$ and $S_y$ in the XYZ model, which does not matter for tetragonal lattice symmetry.

\subsubsection{Spin-Flop Transition}\label{spinflopsection}
 
\begin{figure}
	\centering
\includegraphics[width=0.5\textwidth]{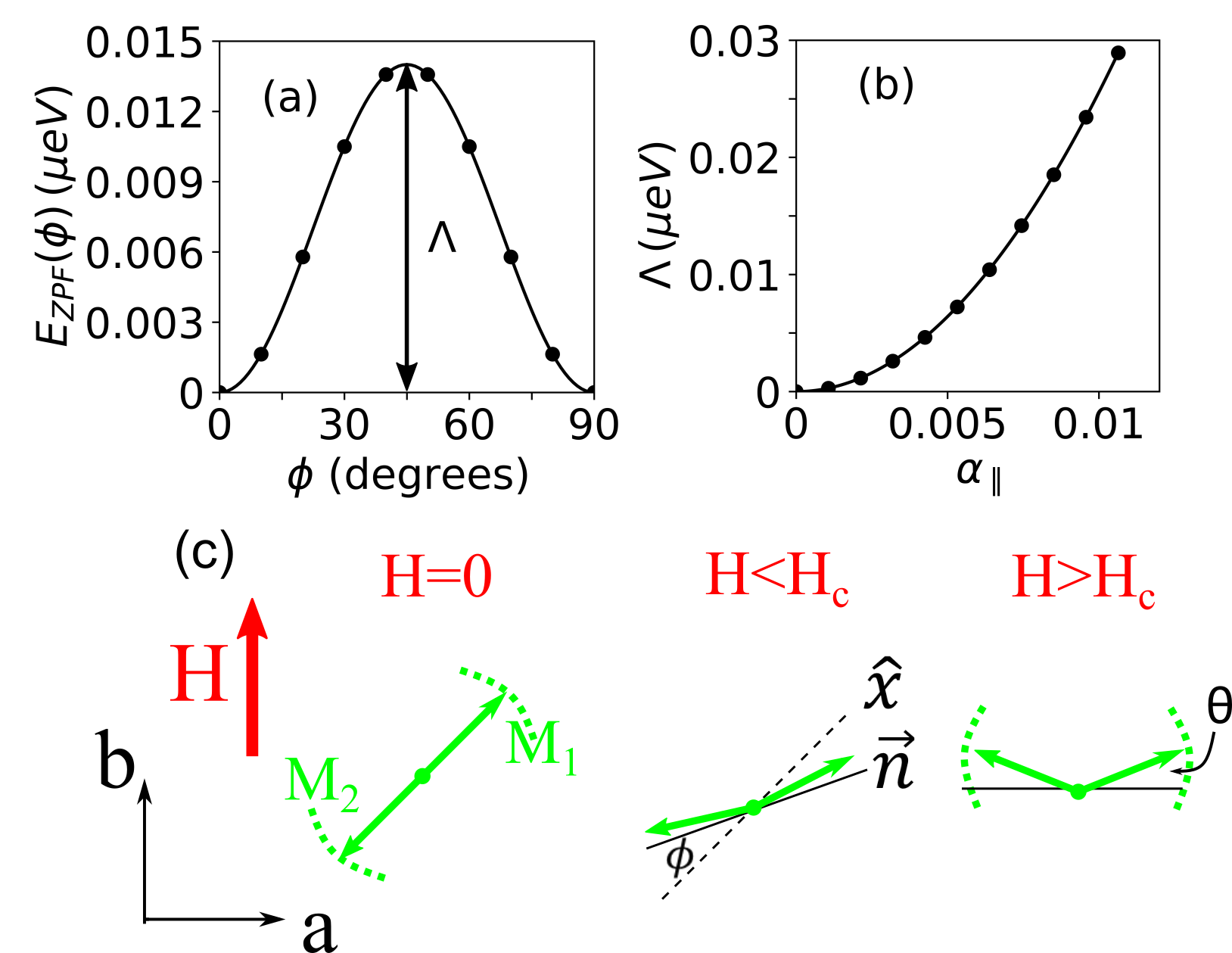}
	\caption{(a) Correction to the ground state energy obtained by summing magnon zero-point fluctuations for different direction of ordered moment, $\phi$. An $\alpha_\parallel=0.0074$ has been used. The peak height, $\Lambda$, is equal to the energy difference between $\phi=0$ (ordered moment along $\hat{x}$) and $\phi=45^\circ$ (ordered moment along $a$ )(b) $\Lambda$ as a function of the exchange anisotropy, $\alpha_\parallel$ (c) Configurations of the two sub-lattice magnetizations ,$\vec{M}_1$ and $\vec{M}_2$, at H=0, H$<\mathrm{H_c}$ and H$>\mathrm{H_c}$, where $\mathrm{H_c}$ is the critical field of the spin-flop transition. The canting between $\vec{M}_1$ and $\vec{M}_2$ is given by $\theta$. The angle between the staggered magnetization, $\vec{n}=\vec{M}_1-\vec{M}_2$, and the $\hat{x}$ axis is denoted by $\phi$.}
	\label{spinflop}	
\end{figure}

We now show that a bulk MAE given by Eq.~\eqref{QZPF} provides a natural explanation for the observed spin flop transition. At zero field, the sub-lattice magnetizations, $\vec{M}_1$ and $\vec{M}_2$, are anti-parallel and lie along $\hat{x}$ or $\hat{y}$ preferred by $E_\mathrm{ZPF}(\phi)$. When a sufficiently large magnetic field is applied along (0, 1, 0), $\vec{M}_1$ and $\vec{M}_2$ are re-oriented almost perpendicular to the field if the Zeeman energy gain by canting $\vec{M}_1$ and $\vec{M}_2$ towards the field (denoted by $\theta$ in Fig.~\ref{spinflop}(c)) is sufficient to overcome the MAE generated by zero-point fluctuations. This leads to a spin-flop transition at a finite critical magnetic field, $H_c$, where $g\mu_B H_c \cos(\theta)\sim \Lambda$. 

A spin-flop transition within the $ab$ plane explains the observed change in (1, 0, 0) magnetic Bragg peak intensity as well as the intensity of the in-plane mode shown in Fig.~\ref{SPINSdata}. For a domain where the ordered moment lies along $\hat{x}$ at zero field, its in-plane fluctuation is polarized along $\hat{y}$ (Ordered moment and in-plane spin fluctuation are denoted by the solid arrow and dotted line in Fig.~\ref{spinflop}(c)). Neutron scattering is only sensitive to the spin component along (0, 1, 0), which is $45^\circ$ from $\hat{x}$ and $\hat{y}$, for scattering near $\bf{Q}$ = (1, 0, 0). Therefore, only half of the Bragg peak and in-plane magnon mode intensity at (1, 0, 0) are detected at zero field. When the field along (0, 1, 0) is greater than $H_c$ of the spin flop transition, all spins are re-oriented perpendicular to the field. The ordered moments are now almost parallel to (1, 0, 0) as shown in Fig.~\ref{spinflop}(c) and hence do not contribute to Bragg peak intensity at (1, 0, 0). On the other hand, the in-plane spin fluctuation is now entirely along the (0, 1, 0) direction. This maximizes the intensity of the in-plane mode. Intensity of the in-plane mode for $H>H_c$ should be twice the intensity at zero field, in agreement with the integrated intensity at H=0 and H=1.2~T shown in Fig.\ref{SPINSdata}(b). Lastly, since spin fluctuation along $c$ is independent of spin orientations within $ab$ plane, the out-of-plane mode should be unchanged across the spin flop transition. This is also consistent with our results in Fig.\ref{SPINSdata}(a).
 
Quantitatively, the spin flop transition happens as a result of the competition between the exchange energy, the MAE generated by zero point fluctuation (Eq.~\eqref{QZPF}) and Zeeman energy due to the canting of the magnetic moments. Summing the three contributions gives the following form for the total energy (per magnetic unit cell):
\begin{align}
\begin{split}
E_\mathrm{tot}=&-4S^2(J_1+J_2+J_3-\frac{1}{2}J_4)\cos(2\theta)\\
&-2HS\sin(\theta)\cos(\phi+\frac{\pi}{4})+E_\mathrm{ZPF}(\phi),
\end{split}
\label{Etot}
\end{align}

where $E_\mathrm{ZPF}(\phi)$ is given by Eq.~\eqref{QZPF} and can be approximated as $\Lambda\sin(2\phi)^2$. Assuming $\vec{M}_1$ and $\vec{M}_2$ initially lie along $\hat{x}$ (same results hold if they lie along $\hat{y}$), Eq.~\eqref{Etot} is minimized with respect to $\theta$ and $\phi$ to find the orientations of $\vec{M}_1$ and $\vec{M}_2$ at a finite field. To directly compare with our data, intensity of the (1,~0,~0) magnetic Bragg peak is computed from the square of the projection of $\vec{n}$ along (0,~1,~0), or $\cos^2(\frac{\pi}{4}-\phi)$. As shown by the solid line in Fig.~\ref{SPINSdata}(b), $\Lambda=0.014(2)~\mu$eV, or equivalently, $\alpha_\parallel=0.0074(4)$ describes the observed field dependence of the Bragg peak intensity reasonably well. Spin wave dispersion determined using the complete set of parameters, $\alpha_\parallel=0.0074$, $\alpha_\perp=0.013$, $J_1=$~4.7~meV, $J_2=$~1.1~meV, $J_3$~=~0.5~meV and $J_4$~=~0.36~meV, is shown in Fig.~\ref{zerofield}. 

To summarize, we have demonstrated the existence of a bulk MAE in the $ab$ plane due to magnon zero-point fluctuations through the observation of a spin-flop transition in the $ab$ plane. This anisotropy also implies that the in-plane magnon mode must acquire a small gap when corrections beyond the linear spin wave approximation are considered. This gap is estimated from the curvature of the semi-classical energy\cite{Rau2018} ($E_\mathrm{MF}+E_\mathrm{ZPF}(\phi)$) to be $4\sqrt{\Lambda J_1\alpha_\perp}\approx 3.5~\mu$eV. Although it is too small to be observed in our experiment or AFMR, the gap may be directly observed in future neutron spin-echo or backscattering experiments. Another check of our model is the direction of ordered moments, which we predict to be 45$^\circ$ from $a$ and $b$. Although this cannot be unambiguously determined from neutron diffraction due to domain averaging, future torque magnetometry\cite{Mirta2010} or angle-dependent susceptibility measurement on a well oriented crystal might be able to provide direct confirmation for our model. 

\section{Discussions}
In the last section, we showed that an XYZ model with an exchange anisotropy of the form, $\alpha_\parallel S_{x,1} S_{x,2}-\alpha_\parallel S_{y,1} S_{y,2}-\alpha_\perp S_{z,1} S_{z,2}$, is a minimal model that explains the magnitude of both the easy-plane anisotropy and the small (classically forbidden) in-plane anisotropy, determined experimentally from the out-of-plane magnon gap and the critical field of an in-plane spin-flop transition, respectively. A natural question is the microscopic origin of this exchange anisotropy. In particular, whether it is consistent with the general theory given by Eq.~\eqref{DMmodel}. In this section, we address this question by considering an alternative model where the exchange anisotropy is given by Eq.~\eqref{DMmodel}.

The symmetry analyses in Appendix A show that Eq.~\eqref{DMmodel} takes a form of
\begin{align}
D_1(S_y^1S_z^2-S_z^1S_y^2)+\frac{D_1^2}{4J_1}(S_x^1S_x^2-S_y^1S_y^2-S_z^1S_z^2)\label{DMBCO}
\end{align}
for the dominant superexchange between $\mathrm{Cu_1}$ and $\mathrm{Cu_2}$ in Bi$_2$CuO$_4$. The second symmetric anisotropic (SA) term alone gives rise to the magnon gap at the magnetic zone center, which allows the only parameter in this model, $D_1$, to be determined. Notably, the SA term in Eq.~\eqref{DMBCO} takes a form quite similar to the exchange anisotropy in our XYZ model given by $\alpha_\parallel S_{1,x} S_{2,x}-\alpha_\parallel S_{1,y} S_{2,y}+\alpha_\perp S_{1,z} S_{2,z}$ (The two are of the same form when $\alpha_\parallel=\alpha_\perp$). Using the value of $D_1$ determined from the out-of-plane magnon gap, the magnitude of the SA term relative to $J_1$ is estimated to be $\frac{D_1^2}{4J_1^2}\sim $ 0.01. Incidentally, this is also on the same order of magnitude as $\alpha_\parallel$ and $\alpha_\perp$ estimated in the last section. It is therefore tempting to conclude that the exchange anisotropy in our XYZ model originate from the SA term in Eq.~\eqref{DMBCO}. On the other hand, the only difference between Eq.~\eqref{DMBCO} and the XYZ model is the existence of an accompanying DM term in Eq.~\eqref{DMBCO} that is an order of magnitude larger. However, as we show below, such a large DM term is inconsistent with the observed magnon spectrum.  

\begin{figure}
	\centering
\includegraphics[width=0.5\textwidth]{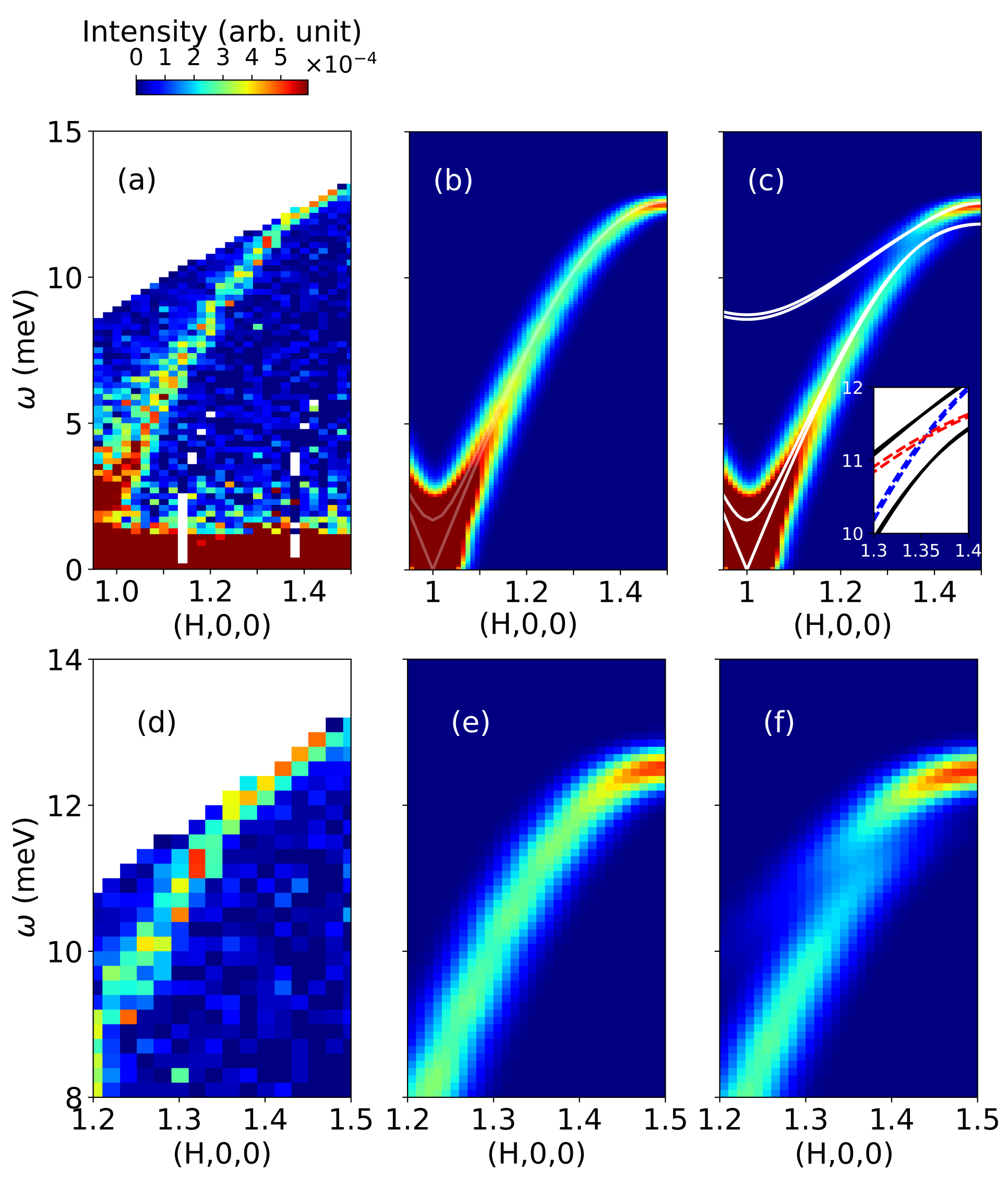}
	\caption{(a,b,c) Comparison between the (a) measured INS spectrum along (0,0,H) (same figure as Fig.~\ref{zerofield}(a)) and the calculated magnon spectrum for (b) XYZ model and (c) DM+SA model. The parameters used in (b) are: $\alpha_\parallel=0.0074$, $\alpha_\perp=0.013$, $J_1=4.7$~meV, $J_2=$~1.1~meV, $J_3$~=~0.5~meV and $J_4$~=~0.36~meV. (c) uses the same parameters for the Heisenberg interactions, $J_1$ to $J_4$, and $D_1=0.22J_1$ for DM+SA interactions between $\mathrm{Cu_1}$ and $\mathrm{Cu_2}$. The calculation is performed using the SpinW package\citep{Toth_2015} taking into account both a Q resolution of $dQ=0.05~\AA^{-1}$ estimated from width of the (1,~0,~0) Bragg peak, and an energy resolution of the form $dE=(7.2\times10^{-4}E^2-5.4\times10^{-2}E+0.97)$~meV calculated for $\mathrm{E_i}=25~$meV at HYSPEC. The white solid lines in (b) and (c) are calculated magnon dispersions. The inset of (c) shows the magnon dispersion of the DM+SA model close to the anti-crossing. The dashed lines are calculated by setting the DM interaction to 0 while keeping the SA term, which removes the anti-crossing. (d,e,f) same as (a,b,c) but zoomed into the region close to the anti-crossing.} 
	\label{DM}	
\end{figure}

In Fig.~\ref{DM}, we compare the measured magnon dispersion along H (Fig.~\ref{DM}(a),(d)) to predictions by the XYZ model (Fig.~\ref{DM}(b), (e)) and the model with DM and SA terms given by Eq.~\eqref{DMBCO} (or the `DM+SA' model, Fig.~\ref{DM}(c), (f)). Clearly, our data is well reproduced by the XYZ model. Although the magnon spectrum within the DM+SA model has an overall shape and intensity similar to the XYZ model, it predicts an anti-crossing close to the zone boundary at around (1.35, 0, 0) and an energy transfer of $\omega\sim11$~meV. This is emphasized in Fig.~\ref{DM}(f), where we zoom into the region close to the anti-crossing. Existence of an anti-crossing in the magnon spectrum can be understood as follows. As we show in Appendix A, although XYZ interaction (and hence the SA term in Eq.~\eqref{DMBCO}) between $\mathrm{Cu_1-Cu_2}$ is identical to that between $\mathrm{Cu_3-Cu_4}$, the sign of the DM interaction is reversed for the two bonds. The translational symmetry by $\it{half}$ of the structural unit cell present in the XYZ model is therefore removed by including the DM interaction. In other words, the magnetic unit cell is now the same size as the structural unit cell, and consists of $four$ rather than $two$ Cu$^{2+}$ ions. Consequently, in addition to the two magnon modes already present when a smaller unit cell is used (blue dashed line in the inset of Fig.~\ref{DM}(c)), one expects two additional magnon modes obtained by folding the zone boundary magnon along (H, 0, 1) to (H, 0, 0) (red dashed line). The DM interaction couples these two sets of magnon modes, and gives rise to an anti-crossing between them that scales with  $D_1$. Although we have used a model where $J_1$ is dominant, and therefore only considered its exchange anisotropy, we emphasize that the above argument is based on symmetry and should be applicable if Eq.~\eqref{DMBCO} is considered for other bonds. For example, since the exchange paths for $J_1$ and $J_3$ have the same symmetries, a large DM interaction along either or both exchange paths is expected to produce an anti-crossing similar to that shown in Fig.~\ref{DM}(c, f).

As shown in Fig.~\ref{DM}(f), the predicted anti-crossing should show up as a small splitting in the magnon dispersion. To detect this subtle feature in our experiment, we used a relatively small E$_i$~=~25~meV which offers a good resolution ($\sim$~0.5~meV) at the energy transfer of the predicted splitting, but partially obscures the top of the magnon dispersion due to kinematic constraint. Even with this limitation, after closely examining our data in Fig.~\ref{DM}(a) and Fig.~\ref{DM}(d), we could not find any sign of splitting in the magnon dispersion. Therefore, even though it is tempting to associate the exchange anisotropy in our XYZ model with the SA term in Eq.~\eqref{DMBCO} as discussed at the beginning of this section, our data does not support the existence of an accompanying DM term that is an order of magnitude larger. This suggests that the spin interactions in Bi$_2$CuO$_4$ might deviate from the general form given by Eq.~\eqref{DMmodel} with a dominant DM term. Deviation from the general theory might be a consequence of the complex geometry of the exchange path in Bi$_2$CuO$_4$. Alternatively, this deviation might be explained by participation of the heavy bismuth ion in mediating the superexchange interaction in Bi$_2$CuO$_4$ as shown by nuclear resonance measurement\cite{Gippius1998}. Inclusion of non-magnetic ions with strong SOC has been shown to dramatically modify the exchange anisotropy in $3d$ transition metal magnets\cite{Stravropoulos2019, Xu2018}. Understanding the effects of bismuth ions on the exchange anisotropy in Bi$_2$CuO$_4$ from first principle should be a focus of future theoretical work.    

\section{Conclusions}   
We have carried out inelastic neutron scattering to study the low energy magnetic excitations in the tetragonal anti-ferromagnet Bi$_2$CuO$_4$. We found a gapless and a gapped magnon mode at low energy that are attributed to spin fluctuations in and out of the easy-plane. Our results resolve the long standing controversies between early INS and AFMR results regarding the low energy excitations in this material, and confirm the latter. By studying the field dependence of (1,~0,~0) magnetic Bragg peak intensity and that of the in-plane mode, we directly observed a spin flop transition in the $ab$ plane at $\sim0.4$~T that was only inferred from previous bulk magnetizations measurements. This indicates the existence of a bulk magnetic anisotropy energy (MAE) in the $ab$ plane which is classically forbidden. We explained all our observations by carrying out spin wave analysis of a minimal anisotropic spin model motivated by the lattice symmetry (which we refer to as an XYZ model). In addition to reproducing the observed magnon dispersion, our model quantitatively explains the critical field of the spin-flop transition, and hence the classically forbidden in-plane MAE through a quantum order by disorder mechanism. In addition to the XYZ model, we also considered an alternative model with a dominant antisymmetric DM interaction, motivated by the conventional theory of exchange anisotropy in cuprates. We found our data is inconsistent with the presence of a large DM interaction, suggesting a departure from the conventional theory in Bi$_2$CuO$_4$ with a complex exchange route.
\section{Acknowledgement}
Work at the University of Toronto was supported by the
Natural Science and Engineering Research Council (NSERC)
of Canada. We acknowledge the support of the National Institute of Standards and Technology, U.S. Department of Commerce, in providing the neutron research facilities used in this work. This research used resources at the Spallation Neutron Source, a DOE Office of Science User Facility operated by the Oak Ridge National Laboratory. Use of the MAD beamline at the McMaster Nuclear Reactor is supported by McMaster University and the Canada Foundation for Innovation.
\section{Appendix}
In this section, we present a symmetry analysis of the general magnetic Hamiltonian in Bi$_2$CuO$_4$, 
\begin{align}
\mathcal{H}=\frac{1}{2}\sum_{ij} \mathcal{H}_{i,j},\label{Hm}
\end{align}
and determine its magnon dispersion in the magnetically ordered state using linear spin wave theory. As discussed in the introduction, the most general pairwise interaction between site $i$ and $j$, $\mathcal{H}_{i,j}$ , is given by
\begin{align}
\mathcal{H}_{i,j}=J_{i,j}\vec{S}_i\cdot\vec{S}_j+\vec{A}_{i,j}\cdot(\vec{S}_i\times\vec{S}_j)+\vec{S}_i^{\mathsf{T}} \mathsf{M}_{i,j}\vec{S}_j.\label{interaction}
\end{align}
The three terms proportional to $J_{i,j}$, the vector $\vec{A}_{i,j}$ and the symmetric matrix $\mathsf{M}_{i,j}$ denote the isotropic Heisenberg interaction, the antisymmetric and symmetric anisotropic exchange interactions, respectively.

\subsection{Symmetry Analysis}\label{Symmetryanalysis}
In this subsection, we present a symmetry analysis of the exchange anisotropy parametrized by the vector $\vec{A}$ and the symmetric matrix $\mathsf{M}$ of Eq.~\eqref{interaction} for the dominant superexchange paths, $J_1$ to $J_4$, as well as the bonds related to them by symmetry. Although we only considered the exchange anisotropy for the dominant exchange path, $J_1$, in our spin wave analysis, the symmetry analysis below is provided for all exchange paths for completeness.  Fig.~\ref{structure}(b)-(c) shows examples of relevant local symmetry operations, including centers of inversion at $[\frac{1}{2},\frac{1}{2},\frac{n}{2}]$ (red stars in Fig.~\ref{structure}(b)), two-fold rotations around $\hat{y}$ axes through $[\frac{1}{2},\frac{1}{2},\frac{1}{4}+\frac{n}{2}]$ (yellow ovals in Fig.~\ref{structure}(b)), and four-fold rotations around $c$ axes passing through the chains of Cu-atoms (yellow squares in Fig.~\ref{structure}(c)). 

\subsubsection{$J_1$ and $J_3$}
$J_1$ and $J_3$ in Fig.~\ref{structure}(b) have the same local symmetry characterized by a two-fold axis along $\hat{y}$ through the center of the bond (shown by yellow ovals). Focussing on $J_1$ between $\mathrm{Cu_1-Cu_2}$, the two-fold rotation around $\hat{y}$ maps $S_{1,x}\leftrightarrow -S_{2,x}$, $S_{1,y}\leftrightarrow S_{2,y}$, $S_{1,z}\leftrightarrow -S_{2,z}$. This constrains the vector, $\vec{A}$,  to lie perpendicular to $\hat{y}$, and therefore take a form of $(A_{1,x},0,A_{1,z})$. Within linear spin wave theory, one only needs to consider the component of $\vec{A}$ parallel to the ordered moment. For ordered moment perpendicular to $c$, $\vec{A}$ can be taken as $(A_{1,x},0,0)$ without loss of generality, which justifies the form of Eq.~\eqref{DMBCO} used in the DM+SA model. 

Similarly, the symmetric matrix, $\mathsf{M}$, takes a form of
\begin{align}
\begin{pmatrix}
J_{xx}^1&0&J_{xz}^1\\
0&J_{yy}^1&0\\
J_{xz}^1&0&J_{zz}^1
\end{pmatrix},
\end{align}
where the super-script indicates that the parameters are for the super-exchange path, $J_1$.

Interactions along other symmetry equivalent bonds, such as $\mathrm{Cu_1^\prime-Cu_2}$, in the $ab$-plane can be generated by a four-fold rotation around the vertical axis through $\mathrm{Cu_2}$, which maps $S_{1,x}\leftrightarrow -S_{1^\prime,y}$, $S_{1,y}\leftrightarrow S_{1^\prime,x}$, $S_{1,z}\leftrightarrow S_{1^\prime,z}$ and $S_{2,x}\leftrightarrow -S_{2,y}$, $S_{2,y}\leftrightarrow S_{2,x}$, $S_{2,z}\leftrightarrow S_{2,z}$. Consequently, the vector, $\vec{A}$,  along this bond is given by $(0,-A_{1,x},0)$ and the matrix, $\mathsf{M}$, is given by
\begin{align}
\begin{pmatrix}
J_{yy}^1&0&0\\
0&J_{xx}^1&-J_{xz}^1\\
0&-J_{xz}^1&J_{zz}^1
\end{pmatrix}
\end{align} 
 
Symmetry equivalent interactions along the $c$ axis, such as that between $\mathrm{Cu_3-Cu_4}$, can be generated by inversion about the point $[\frac{1}{2},\frac{1}{2},\frac{1}{2}]$. This maps $\vec{S}_1\leftrightarrow \vec{S}_4$ and $\vec{S}_2\leftrightarrow \vec{S}_3$. An antisymmetric term, $\vec{A}\cdot(\vec{S}_1\times\vec{S}_2)$,  between $\mathrm{Cu_1-Cu_2}$ gives $\vec{A}\cdot(\vec{S}_4\times\vec{S}_3)=-\vec{A}\cdot(\vec{S}_3\times\vec{S}_4)$  between $\mathrm{Cu_3-Cu_4}$. In other words, the antisymmetric interaction changes sign when translated by half of the structural unit cell along the $c$ direction. However, the symmetric interaction characterized by $\mathsf{M}$ is left invariant by inversion. We therefore arrive at the following observations. If only the symmetric term (e.g. the XYZ model considered in the main text) is present, the primitive unit cell of the magnetic Hamiltonian is half of the structural unit cell. On the other hand, the full structural unit cell has to be used if one also considers the antisymmetric term.
\subsubsection{$J_2$}
The center of inversion between $\mathrm{Cu_1}$ and $\mathrm{Cu_4^\prime}$ (red star in Fig.~\ref{structure}(b)) implies that $\vec{A}=0$ between them. However, all terms in the symmetric matrix, $\mathsf{M}$, are allowed by symmetry,
\begin{align}
\begin{pmatrix}
J_{yy}^2&J_{xy}^2&J_{xz}^2\\
J_{xy}^2&J_{xx}^2&J_{yz}^2\\
J_{xz}^2&J_{yz}^2&J_{zz}^2
\end{pmatrix}
\end{align}.

Like in the case of $J_1$, other symmetry equivalent interactions in the $ab$ plane are generated by four-fold rotation. Those along the $c$ direction are generated by two-fold rotations. For example, two-fold rotation around the $\hat{y}$ axis through $[\frac{1}{2},\frac{1}{2},\frac{1}{4}]$ maps $\mathrm{Cu_1}\rightarrow \mathrm{Cu_2}$, $\mathrm{Cu_4^\prime}\rightarrow \mathrm{Cu_3}$, and simultaneously changing the sign of $S_x$ and $S_z$, while leaving that of $S_y$ unchanged. The matrix $\mathsf{M}$ for the bond $\mathrm{Cu_3-Cu_2}$ is therefore
\begin{align}
\begin{pmatrix}
J_{yy}^2&-J_{xy}^2&J_{xz}^2\\
-J_{xy}^2&J_{xx}^2&-J_{yz}^2\\
J_{xz}^2&-J_{yz}^1&J_{zz}^2
\end{pmatrix}
\end{align}.

The above analysis shows that certain entries of $\mathsf{M}$ for the super-exchange pathway $J_2$ change sign when translated by half of the structural unit cell along $c$, indicating that the structural unit cell must be used when carrying out spin wave analysis including the full symmetric anisotropy for $J_2$.

\subsubsection{$J_4$}
The anisotropic term between $\mathrm{Cu_1-Cu_3}$ is constrained by the four-fold rotation about the axis passing through them (yellow square in Fig.~\ref{structure}(c)), which maps $S_{1,x}\leftrightarrow -S_{1,y}$, $S_{1,y}\leftrightarrow S_{1,x}$, $S_{1,z}\leftrightarrow S_{1,z}$ (similar relations hold for $\vec{S}_3$). This constrains the vector, $\vec{A}$, to be $(0,0,A_4)$, and the matrix, $\mathsf{M}$, to be
\begin{align}
\begin{pmatrix}
J_{xx}^4&0&0\\
0&J_{xx}^4&0\\
0&0&J_{zz}^4
\end{pmatrix}
\end{align}.

Note that the general symmetry allowed anisotropic terms along this bond do not break the in-plane spin rotational symmetry. The bond $\mathrm{Cu_3^\prime-Cu_1}$ is related to $\mathrm{Cu_1-Cu_3}$ by a two-fold rotation about the $\hat{y}$-axis through $[\frac{1}{2},\frac{1}{2},\frac{1}{4}]$ followed by an inversion about $[\frac{1}{2},\frac{1}{2},0]$, which maps $S_{3,x}\leftrightarrow -S_{1,x}$, $S_{3,y}\leftrightarrow S_{1,y}$, $S_{3,z}\leftrightarrow -S_{1,z}$ and $S_{1,x}\leftrightarrow -S_{3^\prime,x}$, $S_{1,y}\leftrightarrow S_{3^\prime,y}$, $S_{1,z}\leftrightarrow -S_{3^\prime,z}$. This operation changes the sign of $\vec{A}$ while leaving $\mathsf{M}$ unchanged. 
\subsection{Spin Wave Calculation}\label{Calculationdetails}
We first define a locally rotated coordinate system ($x^\prime y^\prime z^\prime$) where $\hat{z}^\prime$ is along the direction of ordered moment. For Cu$^{2+}$ in the first sub-lattice ($\vec{M}_1$ in Fig.~\ref{structure}(c)), the transformation to the new coordinate system is defined by:
\begin{align}
\begin{split}
S_x&=-\sin(\phi)S_{y^\prime}+\cos(\phi)S_{z^\prime}\\
S_y&=\cos(\phi) S_{y^\prime}+\sin(\phi)S_{z^\prime}\\
S_z&=-S_{x^\prime}.
\end{split}\label{Coordinatetransform}
\end{align}. 

Coordinate transformation for the other sub-lattice ($\vec{M}_2$) is obtained by replacing $\phi\rightarrow \pi+\phi$ in Eq.~\eqref{Coordinatetransform}. Spin components in this new coordinate system are then expressed as boson creation/annihilation operators via the Holstein-Primakoff transformation as:
\begin{align}
\begin{split}
S_{z^\prime}&=\frac{1}{2}-a^\dagger a\\
S_{+^\prime}&=a\\
S_{-^\prime}&=a^\dagger.
\end{split}
\end{align}

After a Fourier transformation, the magnetic Hamiltonian given by Eq.~\eqref{Hm} can be written as a quadratic boson Hamiltonian in the momentum space:
\begin{align}
\mathcal{H}=\frac{1}{2}\sum_{\mathbf{Q}}\psi^\dagger \mathsf{H}_{\mathbf{Q}}\psi.
\label{Hk}
\end{align}

For a primitive unit cell with $n$ magnetic ions, $\psi=(a_{1,\mathbf{Q}},..., a_{1,\mathbf{Q}},a^\dagger_{n,-\mathbf{Q}},...,a^\dagger_{n,-\mathbf{Q}})^\mathsf{T}$, and $\mathsf{H}_\mathbf{Q}$ is a $2n\times 2n$ matrix given by\cite{Toth_2015}
\begin{align}
\mathsf{H}_\mathbf{Q}=\begin{pmatrix}
\mathsf{A}_\mathbf{Q}&\mathsf{B}_\mathbf{Q}\\
\mathsf{B}^\dagger_\mathbf{Q}&\bar{\mathsf{A}}_{-\mathbf{Q}}
\end{pmatrix}, 
\label{Matrix}
\end{align}
where $\mathsf{A}$ and $\mathsf{B}$ are $n\times n$ sub-matrices. The eigenvalues can be found by diagonalizing the non-hermitian matrix, $\mathsf{G}\mathsf{H}_\mathbf{Q}$, where $\mathsf{G}$ is a diagonal matrix with the first and last $n$ entries given by 1 and -1, respectively. 
\subsubsection{XYZ model}
A primitive unit cell with 2 Cu ions can be used for the XYZ model. The $2\times 2$ sub-matrices $\mathsf{A}$ and $\mathsf{B}$ from Eq.~\eqref{Matrix} are given by:
\begin{align}
\mathsf{A}_\mathbf{Q}=\begin{pmatrix}
C_{\mathbf{Q}}&E_{\mathbf{Q}}\\
\bar{E}_{\mathbf{Q}}&C_{\mathbf{Q}}
\end{pmatrix},
\label{XYZA}
\end{align}
and
\begin{align}
\mathsf{B}_\mathbf{Q}=\begin{pmatrix}
0&F_{\mathbf{Q}}\\
\bar{F}_{\mathbf{Q}}&0
\end{pmatrix},
\label{XYZB}
\end{align} 

respectively. The parameters, $C_{\mathbf{Q}}, E_{\mathbf{Q}}, F_{\mathbf{Q}}$ have been defined in Eq.~\eqref{XYZparameters}. The eigenvalues can be found analytically in this case\cite{Kowalska1966}, and the results are given by Eq.~\eqref{XYZ}.
\subsubsection{DM+SA model}
As discussed in the last sub-section, a primitive magnetic unit cell the same as the structural unit cell has to be used when DM interaction is included for $J_1$. The sub-matrices $\mathsf{A}$ and $\mathsf{B}$ in this case are given by
\begin{align}
\mathsf{A}_\mathbf{Q}=\begin{pmatrix}
\mathcal{C}_{\mathbf{Q}}&\mathcal{A}_{\mathbf{Q}}&\mathcal{B}_{\mathbf{Q}}&0\\
\bar{\mathcal{A}}_{\mathbf{Q}}&\mathcal{C}_{\mathbf{Q}}&0&\mathcal{B}_{\mathbf{Q}}\\
\bar{\mathcal{B}}_{\mathbf{Q}}&0&\mathcal{C}_{\mathbf{Q}}&\mathcal{A}_{\mathbf{Q}}\\
0&\bar{\mathcal{B}}_{\mathbf{Q}}&\bar{\mathcal{A}}_{\mathbf{Q}}&\mathcal{C}_{\mathbf{Q}}
\end{pmatrix},
\label{DMA}
\end{align}
and
\footnotesize
\begin{align}
\mathsf{B}_\mathbf{Q}=\begin{pmatrix}
0&\mathcal{D}^+_{\mathbf{Q}}+\mathcal{D}^{-}_{\mathbf{Q}}&0&\mathcal{E}_{-\mathbf{Q}}e^{-i2\pi L}\\
\mathcal{D}^+_{-\mathbf{Q}}+\mathcal{D}^{-}_{-\mathbf{Q}}&0&\mathcal{E}_{\mathbf{Q}}&0\\
0&\mathcal{E}_{-\mathbf{Q}}&0&\mathcal{D}^+_{\mathbf{Q}}-\mathcal{D}^{-}_{\mathbf{Q}}\\
\mathcal{E}_{\mathbf{Q}}e^{i2\pi L}&0&\mathcal{D}^+_{-\mathbf{Q}}-\mathcal{D}^{-}_{-\mathbf{Q}}&0
\end{pmatrix}.
\label{DMB}
\end{align}
\normalsize
In Eq.~\eqref{DMA} and Eq.~\eqref{DMB},
\footnotesize
\begin{align}
\begin{split}
\mathcal{A}_{\mathbf{Q}}=&\frac{\delta_1}{4}\biggr[\cos(2\phi)(A_{-\mathbf{Q}}-B_{-\mathbf{Q}})-(A_{-\mathbf{Q}}+B_{-\mathbf{Q}})\biggr]\\ 
\mathcal{B}_{\mathbf{Q}}=&\frac{1}{2}J_4 (1+e^{-i2\pi L})\\
\mathcal{C}_{\mathbf{Q}}=&J_4\cos(\pi L)-J_4+2J_1+2J_2+2J_3\\
\mathcal{D}^+_{\mathbf{Q}}=&\frac{1}{2}\biggr[J_1-\frac{\delta_1}{2}+J_3e^{-i2\pi L}\biggr](A_{-\mathbf{Q}}+B_{-\mathbf{Q}})\\
&-\frac{\delta_1}{4}\cos(2\phi)(A_{-\mathbf{Q}}-B_{-\mathbf{Q}})\\
\mathcal{D}^-_\mathbf{Q}=&\frac{iD_1}{2}(\sin(\phi)F_{-\mathbf{Q}}-\cos(\phi)E_{-\mathbf{Q}})\\
\mathcal{E}_\mathbf{Q}=&\frac{1}{2}J_2(A_{\mathbf{Q}}+B_{\mathbf{Q}})
\end{split},
\end{align}
\normalsize
where $\delta_1=\frac{D_1^2}{4J_1}$, $E_\mathbf{Q}=1-e^{i2\pi(H+K)}$ and $F_\mathbf{Q}=e^{i2\pi H}-e^{i2\pi K}$. Numerical diagonalization using the procedure outlined at the beginning of this sub-section gives the dispersion shown in Fig.~\ref{DM}(c). 

\end{document}